\documentclass[preprint]{emulateapj} 

\shorttitle{5-Planet Nice Model} 
\shortauthors{Batygin Brown Betts} 

\begin{document}
 
\title{Instability-Driven Dynamical Evolution Model of a Primordially 5 Planet Outer Solar System}  
\author{Konstantin Batygin$^1$, Michael E. Brown$^1$ \& Hayden Betts$^2$} 

\affil{$^1$Division of Geological and Planetary Sciences, California Institute of Technology, Pasadena, CA 91125} 
\affil{$^2$Polytechnic School, Pasadena, CA 91106} 

\email{kbatygin@gps.caltech.edu}

\begin{abstract}
Over the last decade, evidence has mounted that the solar system's observed state can be favorably reproduced in the context of an instability-driven dynamical evolution model, such as the ``Nice" model. To date, all successful realizations of instability models have concentrated on evolving the four giant planets onto their current orbits from a more compact configuration. Simultaneously, the possibility of forming and ejecting additional planets has been discussed, but never successfully implemented. Here we show that a large array of 5-planet (2 gas giants + 3 ice giants) multi-resonant initial states can lead to an adequate formation of the outer solar system, featuring an ejection of an ice giant during a phase of instability. Particularly, our simulations demonstrate that the eigenmodes which characterize the outer solar system's secular dynamics can be closely matched with a 5-planet model. Furthermore, provided that the ejection timescale of the extra planet is short, orbital excitation of a primordial cold classical Kuiper belt can also be avoided in this scenario. Thus the solar system is one of many possible outcomes of dynamical relaxation and can originate from a wide variety of initial states. This deems the construction of a unique model of solar system's early dynamical evolution impossible.

\end{abstract}


\section{Introduction}
In the last two decades, discoveries of the Kuiper belt \citep{1993Natur.362..730J}, as well as planets orbiting stars other than the Sun \citep{1995Natur.378..355M}, have supplied the centuries-old quest to understand the formation of the solar system with fresh constraints and insights into physical processes at play. Among a multitude of newly proposed formation scenarios, the ``Nice" model \citep{2005Natur.435..459T, 2005Natur.435..466G, 2005Natur.435..462M} is particularly notable, as it has attained a considerable amount of success in reproducing the various observed features of the solar system. Within the context of the scenario envisioned by the Nice model, giant planets start their post-nebular evolution in a compact, multi-resonant configuration, and following a brief period of dynamical instability, scatter onto their current orbits \citep{2007AJ....134.1790M, 2010ApJ...716.1323B, 2011Natur.475..206W}.

The first success of the Nice model lies in its ability to quantitatively reproduce the observed orbits of the giant planets, as well as their dynamical architecture (i.e. secular eigenmodes of the system) \citep{2005Natur.435..459T, 2009A&A...507.1041M}. Simultaneously, the brief instability, inherent to the model, provides a natural trigger to the Late Heavy Bombardment \citep{2005Natur.435..466G}, as well as a transport mechanism for emplacement of dynamically ``hot" Kuiper belt objects (KBOs) from inside of $\sim 35$AU \citep{2008Icar..196..258L}. Meanwhile, it has been recently demonstrated that survival of a dynamically ``cold" primordial population between Neptune's current 3:2 and 2:1 exterior mean-motion resonances (MMRs) is fully consistent with a Nice model-like evolution of the planets, implying an \textit{in-situ} formation of the cold classical population of the Kuiper belt \citep{2011arXiv1106.0937B}. Finally, the presence of Jupiter's and Neptune's Trojan asteroids has been attributed to chaotic capture of planetesimals during the instability \citep{2005Natur.435..462M, 2007AJ....133.1962N}. 

\begin{figure*}[t]
\includegraphics[width=1\textwidth]{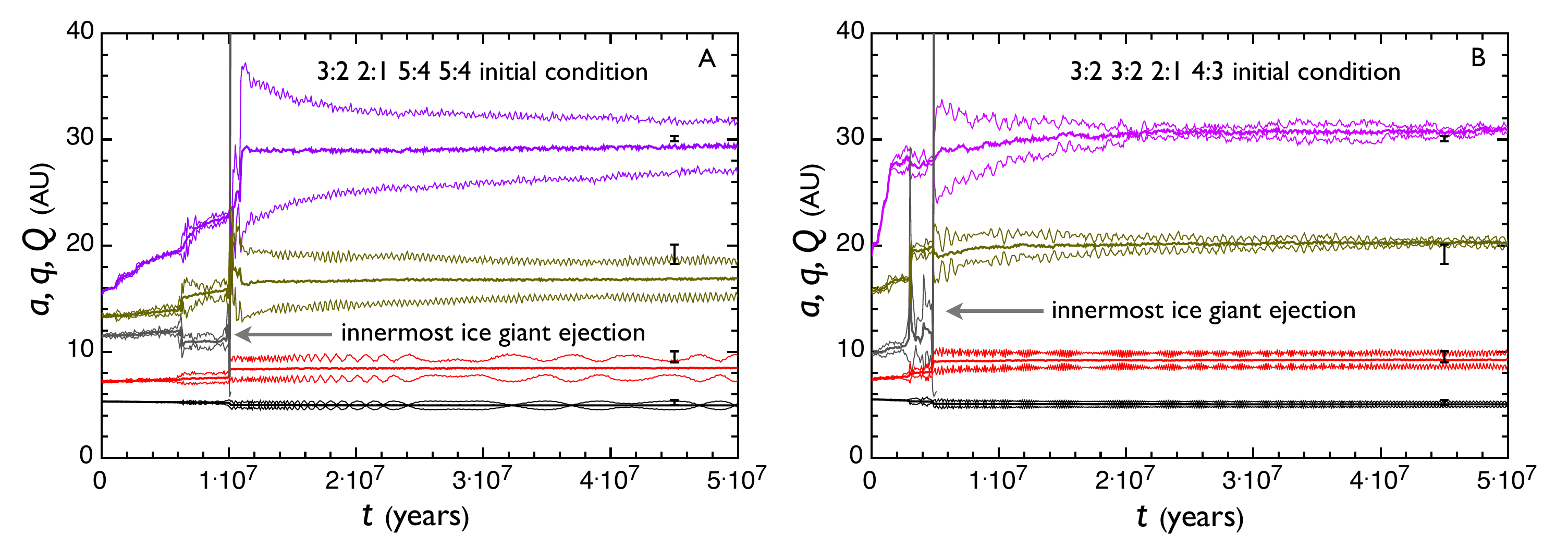}
\caption{Orbital evolution of planets. Each planet's semi-major axis, as well as perihelion and apohelion distances are shown as functions of time. The actual perihelion and apohelion distances of the planets are also shown for comparison as black error bars. In both cases, the innermost ice-giant ejects during the transient phase of instability, leaving behind 4 planets, whose orbits resemble that of the solar system. See main text for a description of the initial conditions.} 
\end{figure*}

All successful realizations of the Nice model to date have been comprised exclusively of the four currently present giant planets. However, there exists no strong evidence that suggests that additional planets were not present in the solar system, at the epoch of the dispersion of the nebula. In fact, theoretical arguments, presented by \citet{2004ApJ...614..497G} point to a possibility of initially forming as many as five ice giants, three of which get subsequently removed via ejections (see however \citet{2007Icar..189..196L}). The dynamical sensibility of such a scenario is further strengthened by the fact that a considerable fraction of standard Nice model simulations result in an ejection of an ice-giant after an encounter with at least one of the gas giants. 

In this paper, we explore an instability-driven dynamical evolution of a 5-planet system (2 gas giants + 3 ice giants) with an eye towards identifying a pathway towards reproduction of solar system-like dynamical architecture. In principle, the realm of possibility available to this study is enormous. Consequently, rather than performing a comprehensive parameter-search, here we limit ourselves to systems that contain an additional Uranus-like planet, with the aim of presenting a few proof-of-concept numerical experiments. The plan of the paper is as follows. In section 2, we describe our numerical setup. In section 3, we show that in our model, the planetary orbits, their secular eigenmodes, as well as various populations of the Kuiper belt are approximately reproduced. We conclude and discuss our results in section 4. 

\section{Numerical Experiments}

The numerical setup of the simulations performed here was qualitatively similar to those presented by \citet{2010ApJ...716.1323B} and \citet{2011arXiv1106.0937B}. Particularly, the five giant planets were initialized in a compact, multi-resonant initial condition, surrounded by a massive planetesimal disk that extended between its immediate stability boundary and $30$AU. 

Two of the three ice-giants were taken to have the same mass as Uranus and were initially placed next to Saturn and as the outermost planet respectively. The middle ice-giant was taken to have Neptune's mass. In all simulations, Jupiter and Saturn started out in a 3:2 MMR, in accord with the results of hydrodynamical simulations of convergent migration of the planets in the solar nebula \citep{2001MNRAS.320L..55M, 2007Icar..191..158M, 2008A&A...482..333P}. The ice giants were also sequentially assembled into first order MMRs by applying dissipative forces, designed to mimic the presence of the nebula \citep{2002ApJ...567..596L}. Following resonant locking, each assembled multi-resonant initial condition was evolved in isolation for 10Myr, as an immediate test of orbital stability. This procedure yielded a total of 81 stable multi-resonant initial conditions. 

The search for adequate dynamical evolutions was performed in two steps. First, we evolved 10 permutations of each initial condition, with planetesimal disks composed of $N=1000$ planetesimals. Disk masses were chosen randomly between $M_{\rm{disk}}^{\rm{min}} = 25M_{\oplus}$ and $M_{\rm{disk}}^{\rm{max}} = 100M_{\oplus}$. The density profiles followed a power-law distribution, $\Sigma \propto r^k$ where the power-law index, $k$, was chosen randomly between $k_{\rm{min}}= 1$ and $k_{\rm{max}} = 2$. The planetesimals were initialized on near-circular orbits ($e \sim \sin i \sim 10^{-3}$). To reduce the already substantial computational cost, self-gravity of the planetesimal swarm was neglected.

Subsequently, we eliminated all initial conditions that did not yield any final systems that were comprised of 4 planets, reducing the number of viable initial conditions to 25. Then, an additional 30 permutations of these initial conditions were integrated with disks composed of $N=3000$ planetesimals (but otherwise identical to those described above). Each integration was performed using the \textit{mercury6} integration software package \citep{1999MNRAS.304..793C} and spanned 50Myr\footnote{Note that here, we make not attempt to time the onset of instability with the late heavy bombardment.}. The calculations were performed on Caltech's \textit{PANGU} super-computer.

After their completion, simulations that were deemed successful were reintegrated with the use of tracer simulations (see \citet{2008Icar..196..258L, 2011arXiv1106.0937B}), to address the dynamical evolution of a locally formed population of KBOs. In particular, each run was supplemented with an additional disk of mass-less particles that resided in the cold classical region of the Kuiper belt (i.e. between the final exterior 3:2 and 2:1 MMRs of Neptune).

\section{Results}

\begin{figure*}[t]
\includegraphics[width=1\textwidth]{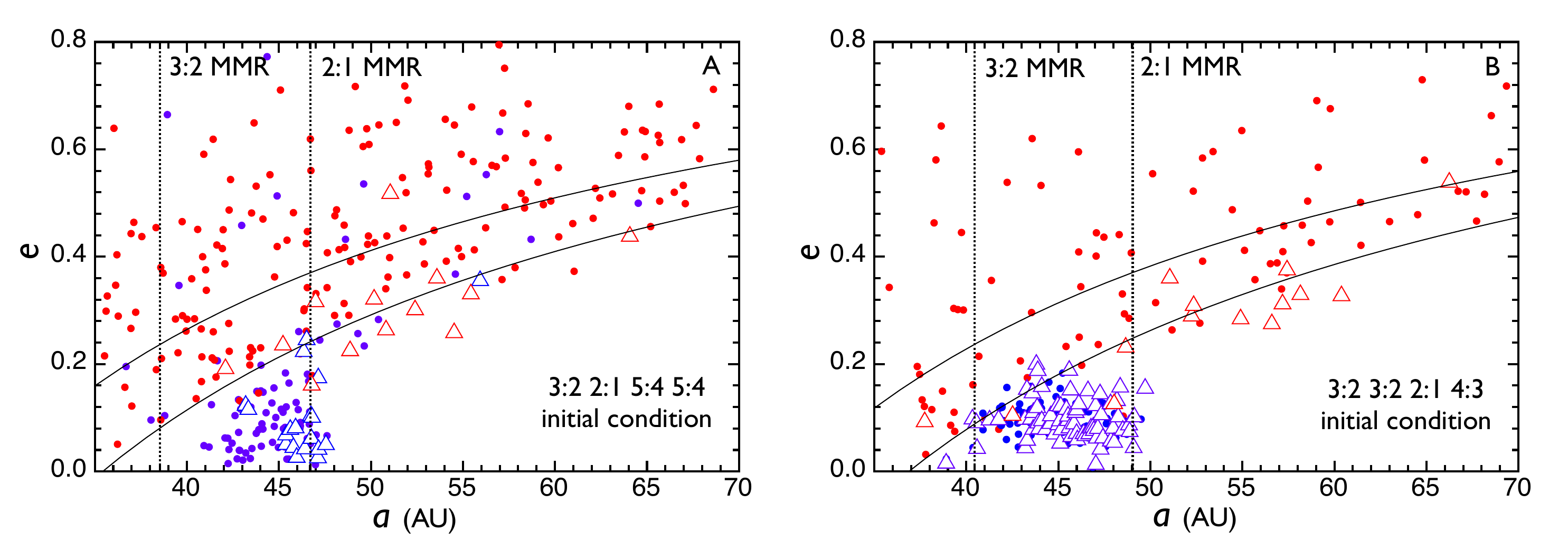}
\caption{Eccentricity distribution of the remnant planetesimal disk. Red dots represent objects that have been dynamically emplaced, while the blue dots depict the locally formed cold classical belt at $t=50$Myr. The red and blue triangles show objects whose orbits are stable on a $500$Myr timescale. The scattered disk is shown with two solid curves and Neptune's exterior 3:2 and 2:1 MMR's are labeled with dashed lines. In the simulation presented in panel A of Figure (1), the cold belt suffers numerous encounters with the ejecting ice-giant, yielding considerable orbital excitation. Moreover, the inner cold belt is further dynamically depleted over $500$Myr of evolution. In the simulation presented in panel B of Figure (1), there is only a single close encounter between the cold belt and the ejecting ice-giant, yielding a dynamically cold orbital structure.} 
\end{figure*}

Out of the 810 integrations that were initially performed, 214 ($\sim 25\%$) cases featured an ejection of a single ice-giant, yielding a system composed of 4 planets. Perhaps unsurprisingly, in most cases the ejected planet is the ice-giant that neighbors Saturn. Of the 750 simulations that were performed following the elimination of initial conditions, 33 evolutions resulted in orbits reminiscent of the solar system. Specifically, we searched for solutions where the Saturn-Jupiter period ratio exceed $2$, while the final semi-major axes of the ice-giants were within $3$AU of their observed counterparts. No strong requirements were placed on the planetary eccentricities and inclinations. 

It is noteworthy that evolving the giant planets onto solar system-like orbits is insufficient for a simulation to be deemed successful for indeed, there are additional constraints that must be satisfied. The first orbital constraint is the reproduction of the secular architecture of the system. The secular orbital angular momentum exchange (i.e. eccentricity evolution) of a planetary system containing $N$ secondaries can be approximately represented as a superposition of $N$ eigenmodes, each corresponding to a fundamental frequency of the system (see \citet{1999ssd..book.....M, 2009A&A...507.1041M}). Physically, the maximum eccentricity that a given planet attains in its secular cycle is equal to the sum of all of its corresponding eigenmode amplitudes. In the context of the traditional Nice model, some difficulty has been noted in correctly reproducing the dynamical character of Jupiter's and Saturn's eccentricity evolution. Particularly, it has been shown that smooth passage of Jupiter and Saturn through the 2:1 MMR has a tendency to under-excite the $g_5$ eccentricity eigenmode as well as their mutual inclinations \citep{2009A&A...507.1041M}. This difficulty can be overcome sometimes by invoking a close-encounter between an ice-giant and the gas giants. Consequently, we have checked, using Fourier analysis (see for example \citet{2009A&A...507.1041M}), the relative strength of the $g_5$ and $g_6$ eigenmodes in all simulations whose orbital end-state resembled the solar system. We did not restrict the success criteria of our simulations to include the correct reproduction of the mean eccentricities of the planets and in some cases, the mean final eccentricities (namely those of the ice-giants) exceeded their observed counter-parts by as much as a factor of $\sim 2$. This is, however, likely an artifact of the coarse representation of the planetesimal disk and the resulting dynamical friction, that we employed in our calculations and should not be viewed as a major drawback.

 In total, we found ten cases (corresponding to eight different initial conditions) where their amplitudes are satisfactorily reproduced. Specifically, in these ten cases, the amplitude of Jupiter's $g_5$ eigenmode exceeds that of the $g_6$ eigenmode, while the amplitudes are roughly the same for Saturn. We made no attempt at quantitatively matching the pair's inclination eigenmodes, however their reproduction does not appear to be problematic \citep{2009A&A...507.1041M}. We also examined the amplitudes of the secular eigenmodes of Uranus and Neptune. Generally it appears that the dynamical architecture of the ice-giants is set in an essentially random manner, depending on the particular encounter history. Consequently, we decided to not use ice-giant secular architecture as a distinctive property in our analysis. 

Successful formation of the Kuiper belt is another important constraint of the Nice model. \citet{2008Icar..196..258L} have shown that the excited populations of the Kuiper belt are naturally emplaced from inside of $\sim 35$AU during the instability. Given that this aspect of the planetary evolution is not particularly different between the 4-planet and the 5-planet scenarios, there is little reason to speculate that the dynamical pathway for formation of the resonant, scattered, and hot classical populations of the Kuiper belt will be inhibited. The same is likely to be true for chaotic capture of Trojan asteroids. The cold classical population of the Kuiper Belt, however, is a different story. 

A series of observational dissimilarities between the cold classical population and the rest of the Kuiper belt (e.g. uniquely red colors \citep{2002ApJ...566L.125T, 2005EM&P...97..107L}; strongly enhanced wide binary fraction \citep{2006AJ....131.1142S, 2010ApJ...722L.204P}) suggest that the cold classicals formed in situ, and maintained dynamical coherence despite Neptune's temporary acquisition of high eccentricity and inclination (a characteristic orbital feature of the cold classical population is inclination that does not exceed $\sim 5^{\circ}$ \citep{2001Icar..151..190B}). In a recent study, \citet{2011arXiv1106.0937B} showed that local formation of the cold classicals is fully consistent with an instability-driven evolution of the planets, given favorable conditions during the instability. Particularly, \citet{2011arXiv1106.0937B} required the apsidal precession and nodal recession rates of Neptune to be comparatively fast to prevent secular excitation of the cold classical orbits, in addition to a sufficiently small apohelion distance of Neptune, to avoid orbital excitation due to close encounters. 

\begin{figure*}[t]
\includegraphics[width=1\textwidth]{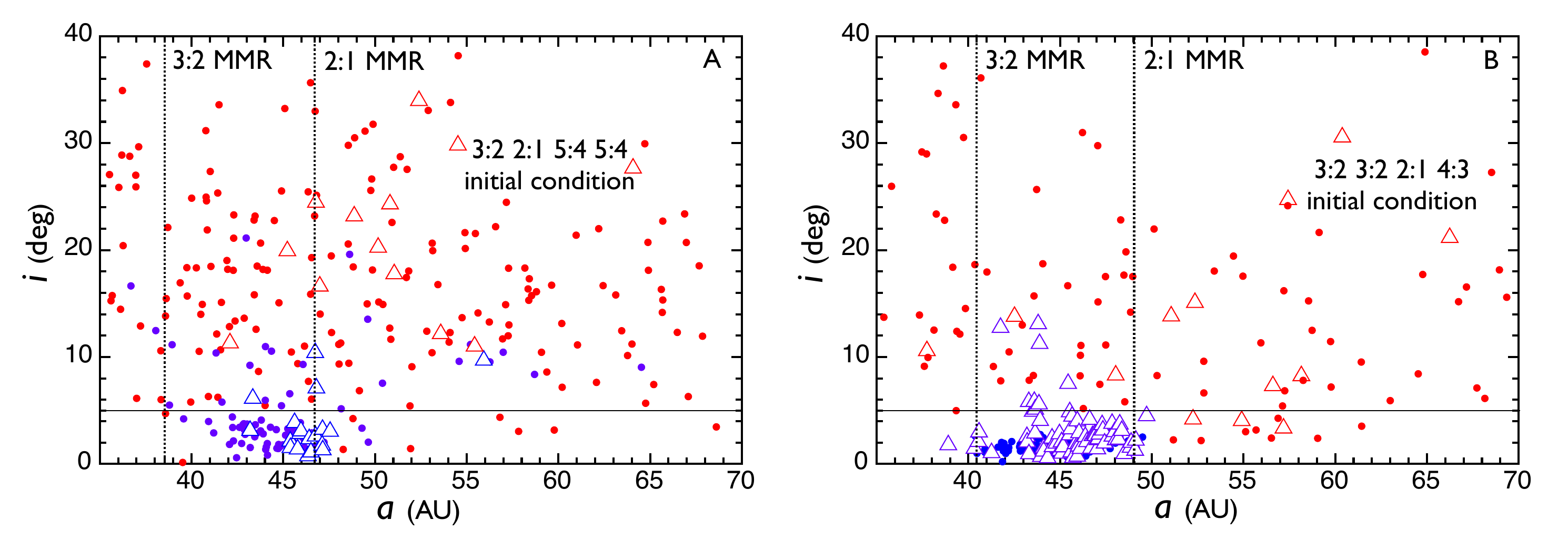}
\caption{Inclination distribution of the remnant planetesimal disk. Red dots represent objects that have been dynamically emplaced, while the blue dots depict the locally formed cold classical belt at $t=50$Myr. The red and blue triangles show objects whose orbits are stable on a $500$Myr timescale. The tentative $i=5^{\circ}$ boundary between the cold and hot classical belts is shown with a solid line, while Neptune's exterior 3:2 and 2:1 MMR's are labeled with dashed lines. In the simulation presented in panel A of Figure (1), the cold belt suffers numerous encounters with the ejecting ice-giant, yielding considerable orbital excitation. Moreover, the inner cold belt is further dynamically depleted over $500$Myr of evolution. In the simulation presented in panel B of Figure (1), there is only a single close encounter between the cold belt and the ejecting ice-giant, yielding a dynamically cold orbital structure.} 
\end{figure*}

In a 5-planet scenario, the retention of unexcited orbits of the cold population can be jeopardized by the ejecting planet. This possibility served as a premise for recalculation of the dynamical evolution of test particles in the cold classical region, with the aid of tracer simulations. We found that in only three of the ten simulations where the secular eigenmodes were successfully reproduced, a primordial cold classical population of the Kuiper belt also retained unexcited orbits. Consequently, it appears that the retention of an unexcited cold belt is not the norm of the 5-planet scenario. This is not surprising, since dynamical excitation can only be avoided if ejection and close encounter timescales are sufficiently short. From a quantitative point of view, our simulations imply that dynamically cold orbits can only be sustained if the ejecting ice-giant spends $\sim 10^4$ years or less crossing the classical Kuiper belt region. As a result, we can expect that the incorporation of yet additional ice-giants into the primordial solar system (as suggested by \citet{2004ApJ...614..497G}) will further diminish the chances of reproducing the cold classical Kuiper belt, since all ejecting planets would have to do so very rapidly.

Two of the three successful simulations discussed above are presented in Figures (1-3). Specifically, Figure (1) shows the orbital evolutions of the runs, while Figures (2) and (3) show the eccentricity and inclination distributions of planetesimals in the Kuiper belt region respectively. The red and blue dots, shown in Figures (2) and (3) depict the orbital distribution of emplaced and local planetesimal populations respectively at $t=50$Myr. The red and blue triangles depict the planetesimals whose orbits remain stable at $t=500$Myr.

The starting multi-resonant initial condition of the simulation presented in panels (A) of the figures is one where Saturn and the first ice-giant are locked in a 2:1 MMR, while both pairs of ice-giants are locked in 5:4 MMRs. The planetesimal disk in this simulation was comprised of $N=3000$ particles and contained a total of 26$M_{\oplus}$. In this evolution, the local population of test particles suffers numerous short close encounters with the ejecting ice giant, yielding a more excited and depleted cold classical population, compared to that of run (B). Note also that at $t = 50$Myr, the ice-giant eccentricities are considerably greater than that of Uranus and Neptune. These high eccentricities do not get damped away by dynamical friction in the following 500Myr of dormant evolution. Consequently, in this simulation, the inner edge of the cold belt gets dynamically depleted over the following 500Myr.  It is furthermore noteworthy that another simulation that originated from the same initial condition reproduced the eigenmodes of the system correctly, although the primordial cold Kuiper belt in this integration was entirely destroyed by close encounters.

The starting multi-resonant initial condition of the simulation presented in panels (B) of the figures is one where Saturn and the first ice-giant are locked in a 3:2 MMR, the inner pair of ice-giants is locked in a 2:1 MMR, and the outer pair of the ice-giants is locked in a 4:3 MMR. In this simulation, the disk consisted of $N=1000$ particles and had a cumulative mass of $42M_{\oplus}$. Incidentally, the frequency spectrum of the eccentricity vectors of Jupiter and Saturn produced in this simulation, matches that of the real Jupiter and Saturn exceptionally well, signaling a nearly ideal reproduction of the secular eigenmodes. Particularly, the simulation yields (in the notation of \citep{1999ssd..book.....M}) $e_{55}^{(sim)}/e_{56}^{(sim)} = 2.28$ and $e_{65}^{(sim)}/e_{66}^{(sim)} = 0.51$ where as the solar system is characterized by $e_{55}/e_{56} = 2.81$ and $e_{65}/e_{66} = 0.68$. The scaling of the eigen-vectors are also well reproduced: $e_{55}^{sim} = 0.0465, e_{66}^{sim} = 0.067$ while the solar system has $e_{55} = 0.0442, e_{66} = 0.0482$ \citep{2009A&A...507.1041M}. In this simulation, the cold Kuiper belt suffers only a single short encounter with the escaping ice-giant, allowing for the orbits (inclinations in particular) to remain dynamically cold.

\section{Discussion}

In this paper, we have presented a successful realization of the Nice model which starts out with 5 planets. The numerical experiments presented here explicitly show that such an evolution is plausible since the resulting 4-planet systems can closely resemble the solar system. Particularly, in both simulations presented here, the secular architecture of the outer solar system is well reproduced. Furthermore the demonstrated survival of a local, primordial cold classical Kuiper belt suggests that all constraints that can be matched with a 4-planet model can also be matched with a 5-planet model to an equal degree of satisfaction. 

It is noteworthy that ejection was not always necessary in our simulations to generate a 4 planet system. In a handful of runs (one of which successfully reproduced the secular eigenmodes, but not the cold Kuiper belt), one of the ice-giants ended up merging with Saturn. In principle, such a scenario may help explain Saturn's enhanced metallicity in comparison with Jupiter. Although, here again the explanation is not unique (see \citet{1982P&SS...30..755S} and the references therein).

In a traditional realization of the Nice model, the rate of successful reproduction of the secular eigenmodes is rather low i.e. $\sim 10\%$ of the integrations for a favorable initial condition \citep{2010ApJ...716.1323B}. This is in part because an ice-giant/gas giant encounter often leads to an ejection of the ice-giant, leaving behind only three planets. Thus, the need for an ice-giant/gas-giant encounter in the orbital history of the solar system is in itself motivation for a 5-planet model. 

The statistics of simulations presented in this work suggest that a 5-planet model is neither more nor less advantageous. Recall that the probability of ending up with only 4 planets is $214/810 \sim 25\%$. The probability of reproducing the secular eigenmodes of Jupiter and Saturn is $10/750 \sim 1.5 \% $. Naively, this yields an overall probability of success of only $\sim 0.4 \%$. However, it is important to keep in mind that the characteristic outcomes are generally dependent on initial conditions\footnote{Interestingly, we do not observe any correlation between the degree to which a given simulation is successful and disk mass.} and runs that originated from the initial condition presented in panels (A) correctly reproduced the secular eigenmodes in 2 out of 30 simulations. This statistic is similar to the 4-planet model. Finally, the $1/30$ probability of also retaining an unexcited cold classical Kuiper belt puts the 5-planet model and the 4-planet model on equal footing in terms of success rate \citep{2011arXiv1106.0937B}. That said, it is important to note that this success rate is only characteristic of the particular 5-planet model that we have constructed. In other words, it is likely that if one allows the mass of the ejected planet to also be a variable parameter, tuning of the initial state may in principle lead to a more frequent reproduction of the solar system.

The results presented in this work imply that the solar system is one of many possible outcomes of dynamical evolution, and can originate from many possible initial conditions. As a result, the possibility of having an extra planet initially present in the system, yet its ejection leaving no observable signature erases any hope for construction of a deterministic model for solar system evolution. The forward process-like nature of the Nice model is not surprising, given that the solar system exhibits large-scale chaos, characterized by Lypunov times that are comparable to orbital timescales, during the instability. Moreover, the similarity between the orbital architectures of simulations whose outcomes were deemed unsuccessful in this work and those of extra-solar planetary systems further confirms that planet-planet scattering is likely to be the physical process responsible for shaping the orbital distribution of planets \citep{2008ApJ...686..603J}. Consequently, we conclude that an instability-driven dynamical history remains a sensible choice as a baseline scenario for solar system's early dynamical evolution. \\

\textbf{Acknowledgments} We thank Alessandro Morbidelli, Hal Levison, David Nesvorny and Peter Goldreich for useful conversations. We thank Naveed Near-Ansari for operational help with the \textit{PANGU} supercomputer. K. Batygin acknowledges supported from NASA's NESSF graduate fellowship.

\end{document}